# Technology Capacity-Building Strategies For Increasing Participation & Persistence In The STEM Workforce

K. M. Moorning

Department of Computer Information Systems, Medgar Evers College of The City University of New York, Brooklyn, NY, USA
kimm@mec.cuny.edu

## ABSTRACT

*This research model uses an emancipatory approach to address challenges of equity in the science, technology, engineering, and math (STEM) workforce. Serious concerns about low minority participation call for a rigorous evaluation of new pedagogical methods that effectively prepares underrepresented groups for the increasingly digital world. The inability to achieve STEM workforce diversity goals is attributed to the failure of the academic pipeline to maintain a steady flow of underrepresented minority students. Formal curriculum frequently results in under-preparedness and a professional practices gap. Exacerbating lower performance are fragile communities where issues such as poverty, single-parent homes, incarceration, abuse, and homelessness disengage residents. Since data shows that more minorities have computing and engineering degrees than work in the field* [1]*, this discussions explores how educational institutions can critically examine social and political realities that impede STEM diversity while capturing cultural cues that identify personal barriers amongst underrepresented groups.*

## 1. INTRODUCTION

The United States Department of Education (DoE) is seeking ways technology can provide better educational outcomes for all students (Jones, Fox, & Neugent, 2015; Future Ready Learning: Reimagining the Role of Technology in Education, 2016). The intensity and complexity of STEM disciplines necessitate expanded opportunities for learning beyond formal departmental silos. An interdisciplinary approach teaching students how to recognize, absorb, and apply knowledge about STEM forms the basis of improving efficiency and stimulating innovation. The Committee on Equal Opportunities in Science and Engineering's (CEOSE) report to Congress called for the creation of a "bold, new initiative for broadening participation." They envisioned large-scale centers that would focus on transforming STEM education with immediate and long-term national impact [2].

Using technology across the disciplines is progressive for the other areas of STEM. Science involves the use of the web-based and computer-based research systems for inquiries and discovery. Math involves the use quantitative data and statistical analysis in numerical expressions. Engineering involves the use of devices and tools for project design and development. This research discusses strategies that best serve intergenerational groups for STEM participation. Collaborative research between institutions of higher education and K-12 schools produces the rigor needed for advancing curriculum and progressing STEM ideals [3]. Using a democratized approach to design centers as learning pathways into the STEM workforce, it is one of the most pervasive models which has discursively survived for decades [4]. Through engagement of educators, public and professional learners, STEM experts, advocacy groups, and corporations this solution addresses underrepresentation issues faced by youth, minorities, and females within specific communities of practice.

## 2. THEORETICAL FRAMEWORK

Several socio-behavioral learning theories were used to establish propositions, assumptions, and empirical validity about the efficacy of STEM centers for workforce diversity. With participatory themes at the core, the aim is to reshape modern thinking about learning among specific learning communities. Positivism and interpretivism are two paradigms we use to explore social facts that shape individual action and "achieve an empathetic understanding by seeing the world through the eyes of the participants" [5]. Using an emancipatory theme, these paradigms empower the youth, minorities, and females involved in the social inquiry. The added elements of critical theory and transformative learning theory focus on how underrepresentation subjugates people's experiences and their understandings of the world.

The core tenet of the transformative learning theory is the notion that adults develop ways to understand the world by considering their own experiences [6]. Emancipatory research with critical inquiry and transformative themes produces knowledge that benefits disadvantaged people and empowers research subjects. Often seen in feminist studies, it relies on the principles of openness, participation, accountability, empowerment, and reciprocity [7]. Born out of the motto "nothing about us, without us," it is a political action which moves research into the "hands of the community being researched" [8]. Prior research revealed that "community programs have the potential to play a critical role" for youth during their developmental period [9]. Students involved in out of school programs make contributions to their communities and are more likely to be interested in STEM [10].

## 3. PRIOR STUDIES

The Metcalf study conducted in 2016 was a cross-tabular analysis of the National Science Foundation's data evaluating "those who have earned their highest degrees in the life sciences" by gender, race, and employment field to show attrition rates out of STEM fields. Applying a critical lens to retention and identity issues showed "the importance of intersecting demographic categories to reveal patterns of experience" for groups whose conditions STEM aims to improve. Including the experiences of marginalized groups helps researchers make informed decisions about policy, practice, and change. This capacity-building research applies Metcalf themes to also look for "hidden cues, omissions, and answers to questions unposed to disrupt, destabilize and denaturalize ideologies."

Anon (2017) used participatory research to study women's experiences in STEM from their viewpoints. Using Photovoice participants presented photographs and narratives describing their experiences in STEM fields. Results revealed the importance females place on facilitating positive relationships. Motivational and mentoring strategies for females invoke feelings of success and satisfaction. Some viewed the lack of recognition as limiting their professional effectiveness despite having fostered relationships. Anon suggested that future research investigate how women deal with workplace challenges to understand how gender stereotypes manifest and impact women in male-dominated careers.

A three-year, small-scale targeted STEM workforce "Pipeline to Technology" study led by Professor Kim Moorning as the principal investigator was conducted at Medgar Evers College of The City University of New York in Central Brooklyn, New York. Using an experiential learning model and participatory research design, it produced an evidence-based practicum for increasing STEM participation for undergraduate students. The study evaluated technology training, STEM efficacy, and workforce access for cohorts of minorities and women to forge pathways for them to enter the rapidly expanding NYC technology industry. While looking at STEM preparedness, the results showed that other causal influences like low workforce access affected low participation because the subjects demonstrated technical skill mastery but lacked self-efficacy [10].

Tuft University and the National 4-H Council led a longitudinal study called the "4-H Study of Positive Youth Development" and surveyed more than 7,000 adolescents from diverse backgrounds across 42 U.S. states. Tuft aimed to define, measure, and drive new thinking and approaches to positive youth development around the world. One major conclusion of the study indicated that youth programs must expand and change to address the diverse and changing characteristics, needs, and interests of adolescents and their families. It discovered that structured out-of-school time, leadership experiences, and adult mentoring plays a vital role in helping young people achieve success [11].

## 4. RESEARCH GOALS

This research integrates strategies from the Metcalf, Anon, Moorning, and Tuft studies to support its extracurricular rationale for STEM centers. Using racial and gender classifications from the STEM ecology, we theorize that by identifying cultural and social cues of youth, female, and minority groups, we can create informal learning pathways for increasing STEM participation. Such cues are expected to provide information about how best to fit learning content to learners' situations and are useful in helping educators more easily understand stimulants that increase interest and proficiencies. The association between informal, co-learning activities and STEM motivation allow K-12 schools, colleges, and universities to:

1. Identify and evaluate the issues of equity and access for underrepresented groups and members of fragile communities.
2. Use research and program data to assess the relationships between their minority communities STEM interest, proficiency, and preparedness.
3. Use research and program data to create informal STEM learning spaces that improve STEM proficiency, competency, and preparedness.
4. Apply data-based understandings of STEM performance to improve retention strategies.
5. Use STEM centers as spaces to coordinate with external stakeholders and the broader education community to enhance capacity-building.

## 5. STEM CENTERS

Higher education institutions need collaborative approaches to attract potential STEM candidates. The education pipeline flows into colleges and extends to the workforce. Designing STEM centers as informal learning spaces to engage learners from three communities of practice: pre-college youth, undergraduates, and working professionals is the catalyst for increasing interest, confidence and competency. These centers are specialized labs for developing skills beyond the formal curriculum and closing the professional practices gap. Tables 1 through 3 show the design and purposes for each audience.

**Table 1 - Youth in Stem Lab**

| **Learner Background** | Young minority high schoolers (ages 14 – 17) who need exposure and deep learning in STEM subjects. This fastest growing group of Internet users need critical skills to interpret and be proficient in STEM. |
|---|---|
| **Audience Purpose** | To prepare youth as STEM citizens, and address the academic inequities faced by public school students from fragile communities, we use knowledge-building and motivation techniques to spark their interest, increase their chances of success, and help to reverse some problematic trends. |
| **Proposed Work** | These learners will be engaged in personalized and project-based learning.<br>• Project/App Development (Software and Arduino kits) |

|  | • Competitions, Makerspaces & Challenges (Hackathons)<br>• Technology Expos (similar to science fairs)<br>• Scientific Inquiry (Internet of Things, Artificial Intelligence)<br>• Cyberlearning & Cloud Computing<br>• Analysis & Reasoning.<br>• Peer Collaboration |
|---|---|
| **Learning Purpose** | To design an informal curriculum that builds confidence, interest, and attraction to the STEM majors with skills they will need in high school, college, and beyond. |

**Table 2 – STEM Learning Lab**

| **Learner Background** | Undergraduate female and minority students seeking additional learning credentials for the STEM workforce. |
|---|---|
| **Audience Purpose** | To improve STEM graduation rates at the bachelor's degree levels for women and minorities and close the professional practices gap. |
| **Proposed Work** | Capacity and knowledge-building program that integrates seven key pillars of STEM learning: collaborative problem-solving, computational analysis, project management, agile software design, systems analysis, programming and app development, and data architecture. They will:<br>• Code computer programs using Java, Web design, and database technology<br>• Create project portfolios<br>• Receive career mentoring<br>• Earn industry recognized micro-credentials<br>• Engage in policy discussions |
| **Learning Purpose** | To build self-efficacy, STEM interest, STEM proficiency, and STEM preparedness to increase the number of professionals in the STEM workforce. |

**Table 3 – Workforce Development Lab**

| **Learner Background** | Working professionals who seek persistence in the STEM labor markets through credentialing. |
|---|---|
| **Audience Purpose** | To increase the persistence of working females and minorities in the technology workforce. |
| **Proposed Work** | Capacity and knowledge-building program that integrates seven key pillars of STEM learning: collaborative problem-solving, computational analysis, project management, agile software design, systems analysis, programming and app development, and data architecture. They will:<br>• Code computer programs using Java, Web design, and database technology<br>• Create project portfolios<br>• Receive expert mentoring<br>• Earn industry recognized micro-credentials<br>• Engage in focus group studies<br>• Engage in policy discussions |
| **Learning Purpose** | To build self-efficacy, STEM interest, STEM proficiency, and STEM preparedness to increase the number of professionals in the STEM workforce. |

# 6. RESEARCH DESIGN

This research uses an empirical baseline of information about STEM for youth, females, minorities, and members of fragile communities. Even though information about these sub-groups is already described in education research, capturing personal metadata is crucial for identifying sensitivities, Two stages are used to identify, classify and integrate cultural and social learning cues (CSLC) across three domains: feasibility, institutional outcomes, and project impact. Table 4 lists the analysis and assessment which must be conducted to evaluate the efficacy of STEM centers at the selected institution.

Table 4 – Research Stages & Outcomes

| STAGE 1 | STAGE 2 |
|---|---|
| **FEASIBILITY** | |
| **Identification and classification of cultural and social learning cues (CSLC)** | **Integration of CSLC into the design of informal STEM learning programs** |
| Determine the range of cultural and social issues found in the population.<br>1. How does informal learning address groups of minorities?<br>  • Does the identification of minority groups' learning styles correspond to learning in traditional education?<br>  • How do instructors understand and make use of information about culture for youth, female, and minority learners?<br>2. In what contexts and for what tasks are the cultural identification useful?<br>  • To what extent do STEM-specific tasks interest each group?<br>  • To what extent are instructors building interest in STEM-specific for each group?<br>3. What clues do instructors use to identify STEM learning needs when engaged in technology access activities?<br>4. What aspects of technologies do the learners perceive? | Explore how CSLC can best be utilized in informal STEM learning space for engaging participants engage in tasks to solve problems they will face in their daily lives and the workplace.<br>1. How best to use CSLC metadata in information-access systems?<br>  • To what extent does providing CSLC metadata improve performance and participation?<br>  • Which specific facets of CSLC improve performance most?<br>  • Can CSLC metadata be used to inform other aspects of STEM curriculum?<br>2. How best to correlate CSLC metadata to underrepresentation and the STEM workforce?<br>  • How does CSLC metadata influence activities (project development, peer collaboration, expert mentoring or internships)?<br>  • What level of granularity of CSLC metadata improves workforce skills? |
| **INSTITUTIONAL OUTCOMES** | |
| **Stage 1 Outcomes:**<br>1. An inductive classification of STEM tasks to be used by our target community<br>2. A documented process for designing informally situated learning curriculum<br>3. A collection of project-based activities and associated tasks assigned for each group of learners<br>4. A profile and set of specifications for group metadata<br>5. Cultural barriers and issues | **Stage 2 Outcomes:**<br>1. A customized informal STEM learning model<br>2. A pathway for reducing the professional practices gap<br>3. Spaces for intergenerational STEM learning<br>4. A plan for STEM career development<br>5. A collaborative forum addressing underrepresentation in STEM |

| PROJECT IMPACT | |
|---|---|
| **Stage 1 Assessment Inquiries:**<br>1. What associations exist between informal learning and increasing STEM proficiency for underrepresented groups and members of fragile communities?<br>2. What associations exist between informal learning and increasing STEM persistence for underrepresented groups, and members of fragile communities?<br>3. What personal barriers impede STEM participation for all underrepresented groups?<br>4. What social norms in fragile communities impede STEM participation? | **Stage 2 Assessment Inquiries:**<br>1. What social norms in the STEM workforce impede participation for underrepresented groups?<br>2. What impact does micro-credentialing have on STEM participation for underrepresented groups?<br>3. What impact does peer collaboration have on female STEM confidence?<br>4. What impact do STEM out-of-school programs have on youth STEM interest? |

## 6.1 Theory of Change & Logic Model

Table 5 outlines the theory of change and logic model indicating the resources, inputs, short-term outcomes, and long-term impact for each subset (learners, program, partners). This research's STEM theory of change are based on the following premises:
- Participants involved in a triad of cooperative activities build knowledge and capacity.
- Integrating scientific and technical methodologies increase learners' proficiency and confidence.
- Cyberlearning and program development build learners' proficiency and preparedness.
- Cooperative learning motivates individuals and groups to solve complex problems.
- Competitions and expositions build learners interest and exposure.
- Expert mentorship fosters inclusiveness, persistence, and diversity.
- Micro-credentialing represents skill mastery and influences career choices.

### Table 5 – Theory of Change & Logic Model

| RESOURCES | INPUTS ACTIVITIES | SHORT TERM OUTCOMES | LONG-TERM IMPACT |
|---|---|---|---|
| **LEARNERS** | | | |
| **Minority H.S students** | ▪ Out-of-School Learning<br>▪ Competitions & Tech Events (Hackathons)<br>▪ Peer Learning<br>▪ STEM education path<br>▪ STEM expert mentoring | Increases<br>▪ # of youth in tech<br>▪ # of STEM projects<br>▪ # of STEM majors | Increases<br>▪ STEM interest, knowledge & skills<br>▪ Boosted confidence ratios<br>▪ Boosted proficiency ratios |
| **Female & Minority Undergraduates & Workers** | ▪ Professional Development<br>▪ Collaborative Learning<br>▪ STEM expert mentoring<br>▪ STEM workforce path<br>▪ Career Mentoring<br>▪ Focus Groups | Increases<br>▪ # of STEM projects<br>▪ # of female groups<br>▪ # of STEM job<br>  ○ prepared<br>  ○ access<br>  ○ placements | Increases<br>▪ # of participants in the STEM ecology (connected to STEM expert or workforce)<br>▪ # of STEM professionals with micro-credentials<br>▪ # of candidates prepared for the STEM workforce |
| **PROGRAM** | | | |
| ▪ Out-of-School Program<br>▪ STEM Co-Curriculum<br>▪ STEM Centers | ▪ Tools for Gauging Designing Informal Curriculum<br>▪ Research Centers<br>▪ Communities of Practice<br>▪ Informal STEM Learning | Increases<br>▪ STEM extra-curricular activities<br>▪ STEM co-curricular activities | ▪ Knowledge Building Model STEM learning for increasing proficiency<br>▪ Collaborative Model for developing confidence and competence |
| | • Professional Development | • faculty motivation to design personalized learning objects | • Reduction Model for STEM mitigating professional practices gap<br>• Increased educator capacity for designing STEM learning curriculum<br>• Increased institutional capacity for addressing sensitivities within minority & female groups. |
| **COLLABORATIVE PARTNERS** | | | |
| • **K-12 Administrators**<br>• **STEM Faculty**<br>• **Expert Mentors**<br>• **Professional Coaches**<br>• **STEM Advisory Council**<br>   o **Public Officials**<br>   o **STEM Researchers**<br>   o **Advocacy Groups**<br>   o **Corporations** | • STEM focus groups<br>• STEM policy discussions | Increases<br>• Partnerships<br>   o Schools<br>   o Public officials engaged<br>   o STEM experts<br>   o STEM faculty<br>   o Advocacy groups<br>• Evidence for the research community | Increases<br>• K-12 capacity to advance STEM learning<br>• Capacity for influencing STEM policy through educator-community partnerships<br>• Capacity for workforce diversity through college-corporate partnerships |

## 6.2 Data Management & Evaluation

Evaluating the logic model required capturing basic statistics (descriptive, inferential, frequencies, distributions, correlations). Through formative evaluation, and summative evaluation, the data will expose factors and barriers related to STEM participation. The cognitive, behavioral, and social cues add to the feasibility validity throughout the development process. To conduct the formative evaluation, some instruments are intuitive, and others are available to the educational community. The following data collection is necessary:

- Demographics Survey (age, gender, race, household income, household, marital composition, parents' careers (youth), major (undergraduates) # of years in the STEM (working adults).
- STEM Self-Efficacy Questionnaire (Adults)
- STELAR Pre-College Annual Self-Efficacy Survey (Youth)
- STELAR STEM Career Interest Questionnaire (Adults)
- Student Interest in Technology & Science (Youth)
- STEM Career Knowledge Questionnaire (All)
- Participant Semi-Structured Interviews (All)

For managing the STEM center's data and making strategic decisions, the following data are to be assessed:

- Persistence & Retention Records (Attendance, Project, Work Patterns, Time Patterns)
- The relationship between interest and (household composition, socio-economic status, peer collaboration, mentoring and confidence)
- The relationship between proficiency and (household composition, socio-economic status, peer collaboration, mentoring and confidence)
- The relationship between persistence and (household composition, socio-economic status, peer collaboration, mentoring and confidence)
- Participant Activity Assessment Surveys

Table 6 shows the data inquiries needed for a summative evaluation.

### Table 6 – Data Inquiries

| Data | Data Inquiry |
|---|---|
| **Program Persistence** | Did learners remain engaged with the program over time? What factors (learner and program) appear to be related to persistence in the program? |

| STEM Self-Efficacy | What is the relationship between participation in This research project and changes in each learner's level of STEM Self-Efficacy? |
|---|---|
| **STEM Interest** | What is the relationship between participation in This research and changes in each learner's level of interest in the STEM? |
| **Technology Proficiency** | What is the relationship between participation in This research and changes in each learner's level of proficiency with technology? |
| **Preparedness for STEM Majors (Youth)** | What is the relationship between participation in This research and changes in each youth learner's preparedness for a STEM major? |
| **Preparedness for STEM Careers (Adults)** | What is the relationship between participation in This research and changes in each adult learner's preparedness of a STEM Career? |
| **Expected Deliverables** | Have the expected deliverables been completed and implemented?<br>• Did the project meet it learner-participant goals?<br>• Did the project enlist sufficient expert mentors?<br>• What external agencies participated in the project?<br>• How many relationships with school district partners were maintained?<br>• How many publications in peer-reviewed journals were made?<br>• How many presentations at regional, state, national, or international professional conferences were made? |

The data analysis plan necessary used in this model is outlined in Table 7.

### Table 7 – Data Analysis Plan

| Data Method | Data Analysis | Purpose |
|---|---|---|
| Purposive Sampling | Participant Demographics | Sample minorities and females in New York City locale. This method is expected to improve the generalization performance of the intervention for these groups. |
| Stratified Sampling | Participant Demographics | Draw conclusions from different youth, female, and minority sub-groups. |
| Cross Tabulation | Survey Data | Understand the correlation between different variables collected through survey instruments to show correlations across groups of participants based on patterns, trends, and probabilities within raw data. |
| Propensity score matching | Participant interest<br>Participant job interviews<br>Participant job placement<br>Workforce barriers<br>STEM Job offerings<br>Workforce demographics<br>Workforce Qualifications | Estimate the effect of the intervention by accounting for the covariates (STEM job interviews, job placement, STEM interest) and reduce bias due to confounding variables (workforce barriers, job offerings, professional practices gap) that could impair treatment. |
| Wilcoxon signed-rank tests | Curriculum Outcomes<br>Competitions & Expos<br>Career Mentoring<br>Expert Mentoring<br>Peer Collaboration<br>Cooperative Learning<br>Project Development<br>Scientific Inquiries | Capture the metrics during repeated assessments across interventions and groups of participants. |
| Internal Factor Evaluation | College Setting<br>K-12 participants<br>Adult participants<br>K-12 school partners<br>Instructors Capacity<br>This research Curriculum | Evaluate the strengths and weaknesses of the research partnerships' (college, K-12, instructors, informal curriculum) and project outcomes' (STEM centers). |

|  | STEM Centers |  |
| --- | --- | --- |
| Parametric Tests | STEM Center Intervention STEM Participation | Test the statistical power and detect a significant effect of the intervention on participants. |
| Principal Component Analysis | Participant Attitudes Psychosocial Improvements Behavioral Improvements Improved STEM Proficiencies Female Confidence Changes STEM Workforce Qualifications | Measure principal components to seek internal validity and reduce intervention complexity. |
| Exploratory Factor Analysis | Fragile community traits Participant Intellect Participant Personality Participant Social Attitude | Conduct the multivariate statistical method of multiple regression and partial correlation to postulate the latent variables that underlie patterns in manifest variables (underrepresentation, persistence). |
| Poisson Regression | Number of participants Number of job placements Participant Retention Increase in Interest Increase in Confidence Increase in Proficiency | Measure the effects of the intervention on participants to determine the goodness of fit, confidence limits, likelihoods, and deviances. Perform a comprehensive residual analysis to provide confidence intervals on predicted values. |

## 7. CONCLUSION

The formal college curriculum has proven inadequate in closing the STEM workforce gap. Informal extra-curricular and co-curricular activities support task-oriented and performance-based workforce development.  Campus STEM centers promote inquiry and discovery with long-term, far-reaching implications for transforming practices in out-of-school, afterschool and professional development programs.  Using high-quality workforce training models, peer collaboration, group learning, and mentoring allow participants to gain academic and industry-recognized proficiencies and micro-credentials that build self-efficacy.  In the same way, the participatory research activities create social and behavioral knowledge and tangible beliefs about youth and female communities of practice.  Pre-college open learning spaces expose youth to real-world STEM problems that peak their interest faster at a critical time in their development.  The informal co-curricular design for adult learners builds competency that closes the professional practices gap. This research maps to NSF core values of:
- scientific excellence -- by creating a transformative and innovative learning model;
- organizational excellence -- by developing, motivated, inclusive, and positive workers;
- learning -- by identifying curricular opportunities for professional growth, and sharing our best insights through collaboration;
- inclusiveness -- by embracing contributions from underrepresented groups and fragile communities; and
- accountability for public benefit -- by creating high standards of performance which benefits participants, partners, employers, secondary and post-secondary schools, research agencies, and the public.

Coordinated efforts with external stakeholders contribute to the pertinent dialogue about equity and access challenges. Emancipatory research about minority groups in fragile communities provides a cultural lens not presently addressed in STEM research. Colleges and universities can expect to gain:
- New evidence about fragile communities and STEM underrepresentation.
- Cognitive and non-cognitive data about challenges faced by underrepresented groups.

- Cultural data about acceptable benefits for broadening participation.
- Social and behavioral data about youth, minorities and female communities of practice.
- An informal STEM centers and open space labs for intergenerational student development.
- A micro-credentialing program for closing the professional practices gap.
- A scientific practices model for peaking youth's interest in STEM subjects.
- A greater understanding of STEM pedagogy, curriculum, graduate pathways, and workforce development.

This research has the potential to transform the futures for members of underrepresented groups and fragile communities. Student success factors are based on the high-performance skills learners acquire in achieving dreams for participating and persisting in the STEM workforce. As educators reshape the way they think about lifelong learning along gender, age, and racial lines, the emphasis on establishing propositions and assumptions will provide the empirical validity for redesigning STEM curriculum. In establishing equity and sustainability, it is necessary to influence STEM interest by designing compelling learning activities in learning spaces where skills are mastered without encumbrances. STEM centers are a lifelong learning product where learners can remain engaged for many years. The more significant goal is to support the nuances of a knowledge-building society which encourages society to learn and work smarter. This mechanism advances knowledge across social, cultural, and education domains and provide a clear pathway for increasing the number of minority and female participants persistent in the global STEM workforce.

Professor Kim Moorning is an information technologist, instructional technologist, author, educator, and researcher. With twenty years in higher education at The City University of New York, she teaches in the Department of Computer Information Systems at Medgar Evers College, and coordinates efforts in institutional assessment and accreditation. She is a systems thinker, thought leader, and strategist who promotes innovation in education.  Prior to this, Prof. Moorning worked for Fortune 500 corporations in the banking, brokerage, and legal industries of NYC, and taught business, technology, and professional development courses in the private sector.  She is a graduate of Teachers College - Columbia University, NYU Polytechnic Institute, and Baruch College with emphasis in Computer Information Systems, Management, Instructional Technology, and Education Leadership.   Prof. Moorning also holds certificates in coaching, leadership, digital marketing, and technical training from various academic and business entities.